\documentclass[aps,prd,amssymb,showpacs,floatfix,twocolumn,superscriptaddress,nofootinbib]{revtex4-1}
\usepackage{amsmath,amsfonts,graphics,epsfig,amssymb}
\begin{document}
\title{%
Towards a model of population of astrophysical sources of
ultra-high-energy cosmic rays }
\author{O.E.~Kalashev}
\affiliation{Institute for
Nuclear Research of the Russian Academy of Sciences, 60th October
Anniversary Prospect 7a, Moscow 117312, Russia}
\author{K.V.~Ptitsyna}
\affiliation{Physics Department, M.V.~Lomonosov Moscow State University,
Moscow 119899, Russia}
\affiliation{Institute for Nuclear Research of the Russian Academy of Sciences,
60th October Anniversary Prospect 7a, Moscow 117312, Russia}
\author{S.V.~Troitsky}
\affiliation{Institute for Nuclear Research of the Russian Academy of Sciences,
60th October Anniversary Prospect 7a, Moscow 117312, Russia}


\date{July 11, 2012}

\begin{center}
\begin{abstract}
We construct and discuss a toy model of the population of numerous
non-identical extragalactic sources of ultra-high-energy cosmic rays. In
the model, cosmic-ray particles are accelerated in magnetospheres of
supermassive black holes in galactic nuclei, the key parameter of
acceleration being the black-hole mass. We use astrophysical data on the
redshift-dependent black-hole mass function to describe the population of
these cosmic-ray accelerators, from weak to powerful, and confront the
model with cosmic-ray data.
\end{abstract}
\end{center}
\maketitle

%%%%%%%%%%%%%%%%%%%%%%%%%%%%%%%%%%%%%%%%%%%%%%%%%%%%%%%%%%%%%%%%%%%%%%%

\section{Introduction}
\label{sec:intro}
The origin of ultra-high-energy (UHE) cosmic rays (CRs), that is of cosmic
particles with energies $E \gtrsim 10^{19}$~eV, is presently unknown.
However, there are numerous hints, both in data and in theory, which might
help to constrain possible models of UHECR sources. In particular, the
observation \cite{HiRes-GZK, PAO-GZK, TA-GZK} of the spectrum steepening
consistent with the Greisen--Zatsepin--Kuzmin \cite{G, ZK} (GZK) cutoff,
together with the global isotropy of the arrival directions
(see e.g.\ ~\cite{TA-anisotropy}) and the fact that the UHE
particles are not expected to be confined by the Milky-Way magnetic field
(see e.g.\ \cite{NotConfined}) suggest that the bulk of cosmic rays at
these energies have extragalactic origin\footnote{See however Ref.\
\cite{STdoublet} and references therein where possible exceptions are
discussed.}. Next, the lack of clustering of arrival directions at small
scales is a powerful tool \cite{DubovskyClustering} to constrain the
number density of sources. In recent data of the Pierre Auger Observatory
(PAO), no evidence for clustering at $E \gtrsim 5 \times 10^{19}$~eV is
seen \cite{PAO-clustering} which translates into the number density of $n
\gtrsim 10^{-4}$ sources per cubic Megaparsec, which means that sources of
these extreme particles should not be exceptional and, most probably, some
of them should be located relatively nearby. The latter fact gets further,
though limited by statistical significance, support from the shape of the
GZK feature in the spectrum which does not seem to be very sharp (cf.\
Ref.~\cite{SemikozMild}).

On the other hand, it is a nontrivial task to find particular
astrophysical objects which could serve as UHECR accelerators. Even
without a detailed modelling of the acceleration process, a number of
simple estimates rule out many classes of potential sources. These
simple criteria include in particular the geometrical (Hillas) criterion
\cite{Hillas} and estimates of radiative energy losses of particles being
accelerated (see e.g.\ Refs.~\cite{Aha-losses, Medv-losses}). Analysis of
the modern astrophysical data demonstrates \cite{UFN-HillasPlot} that the
combination of these constraints leaves just a few candidate classes of
sources capable of acceleration of particles to UHE energies. Leaving
aside large-scale structures where interaction losses are expected to
suppress the energy gain, the conventional diffusive (e.g., relativistic
or non-relativistic shock) acceleration may work only in ultrarelativistic
jets, hot spots and lobes of exceptional active galaxies (powerful radio
galaxies and blazars) which are not that abundant in the nearby Universe.
For very special field configurations when synchrotron losses are
suppressed and the curvature radiation dominates, possible acceleration
sites include also gamma-ray bursts (GRBs) and immediate neighbourhood of
supermassive black holes (SMBHs) in the galactic nuclei. While it is
unclear whether these field configurations may be present in GRBs, recent
IceCube results disfavour the GRB scenario anyway \cite{IceCubeGRB} (see
however Ref.~\cite{Dar}). At this level of reasoning, the SMBH environment
remains a viable option.

A natural assumption is that numerous UHECR sources are not
identical -- there should be less and more powerful accelerators where the
maximal energies, injection spectra and fluxes of accelerated particles are
different. Until now, numerous attempts to model the sources of UHECRs and
to confront theoretical predictions with experiments often assumed that
these parameters are fixed once and for the entire Universe (see e.g.\
Refs.~\cite{Berezinsky, Oleg} and numerous other works; see however
Ref.~\cite{Ptuskin} where acceleration in non-identical jets was
considered). While, for numerous sources, the assumption of equal fluxes is
well justifiable (in the sence that only the mean flux of a large sample
of sources is important and this mean flux does not vary significantly
from one region in the Universe to another) and the injection spectrum is
often fixed by the acceleration model, the maximal energies are expected
to vary significantly. As it was recognised in Ref.~\cite{SemikozMaxEn},
these variations affect the observable spectrum seriously. In this work,
we attempt to present a toy model of numerous and different sources of
UHECRs and, within certain assumptions, to confront it with the
experimental data.

To this end, we choose a simple toy model of particle acceleration in the
immediate vicinity of SMBH put forward in Refs.~\cite{NerSem, NerSemTk}.
The reason to choose this particular model is twofold. First, unlike many
other models, it allows~\cite{UFN-HillasPlot} for UHECR acceleration in
numerous nearby sources. Second, as we will see below, within some
realistic assumptions, the acceleration capabilities of a source are
determined by a single parameter, the SMBH mass. At the same time, the
demography of SMBHs is well studied by astrophysicists and we take this
advantage to describe the population of sources easily.

The rest of the paper is organized as follows. In Sec.~\ref{sec:BH}, we
give a brief review of the acceleration model of Refs.~\cite{NerSem,
NerSemTk} and make a bridge between the parameters which determine the
maximal energy of accelerated particles and the SMBH mass. In
Sec.\ref{sec:population}, we discuss the astrophysical data on the SMBH
population, merge them with the acceleration model, calculate the spectrum
of UHECRs with the account of propagation from source to the observer and
compare it with the experimental cosmic-ray data. We obtain a good
agreement with the observed spectrum by fitting the spectrum with only
two continuous parameters, the overall normalization and a single free
parameter of the model. Sec.~\ref{sec:constraints} demonstrates that the
population model of Sec.~\ref{sec:population} satisfies simple
observational constraints: it does not produce too much secondary gamma
rays, it results in an acceptable number density of sources and, with the
best-fit normalization, it does not require enormous luminosity of a
single source. We give our conclusions and discuss our results in
Sec.~\ref{sec:concl}.

\section{A toy model of particle acceleration in the black-hole
magnetosphere}
\label{sec:BH}
A toy model of particle acceleration in the black-hole magnetosphere was
proposed by Neronov et al.~\cite{NerSem, NerSemTk}. Let us briefly discuss
the model and its parameters.

Assume that a stationary rotating black hole without
electric charge is embedded into the external magnetic field, homogeneous
at the horizon distance scale. In general, the magnetic field is inclined
at some angle $\chi$ with respect to the black-hole rotation axis. There is
a well-known exact solution of Maxwell's equations in the Kerr metric for
each inclination angle $\chi$ of an asymptotically homogeneous magnetic
field \cite{BHsolution1, BHsolution2}. For instance, if $\chi=0$, then a
rotation-induced electric field is parallel to the magnetic one on the
symmetry axis and its direction depends on the directions of both the
magnetic field and the black hole's rotation velocity. Thus in the region
near the rotation axis, particles moving along magnetic lines are
accelerated by the electric field.

In this case, radial components of the electric and magnetic fields in
units $\hbar=c=G=1$ in locally non-rotating frame in Boyer-Lindquist
coordinates on the symmetry axis are:
$$
B_{\widehat{r}}=B_{0}\left(1-\frac{4a^2Mr}{(r^2+a^2)^2}\right),
$$
\begin{equation}
\label{eq:1}
E_{\widehat{r}}=-\frac{2aMB_{0}(r^2-a^2)}{(r^2+a^2)^2},
\end{equation}
where $M$ is the black hole's mass, $a \leq M$  is its angular momentum
per unit mass, $r$ is the radial coordinate and $B_{0}$ is the
external homogeneous magnetic field.

Neglecting for the moment the energy losses, the maximal energy gain of the
accelerating particle with a charge $Ze$ is determined by the available
potential difference in the region along the rotation axis,
\begin{equation}
\label{eq:2}
{\cal E}_{\rm max}(a)=\int_{r_{\rm hor}}^{r_{\rm max}} Ze E(r,a)dr,
\end{equation}
$r_{\rm hor}=M+\sqrt{M^2-a^2}$ is the radius of the black-hole horizon
where particle acceleration starts while $r_{\rm max}$ limits the
size of the region along the rotation axis, where acceleration is possible
(in Ref.~\cite{NerSemTk}, it is called ``the vacuum gap'' due to the
absence of numerous charged particles in this region except of the
single test particle being accelerated, which does not change
the electromagnetic field).

Of course, one can rewrite Eq.~(\ref{eq:2}) for the potential difference
and ${\cal E}_{\rm max}$ in terms of the distance-averaged electric field
$\bar{E}$,
$$
{\cal E}_{\rm max}(a)= Ze \bar{E}(a) H ,
$$
$$\bar{E}(a)=\frac{1}{H}\int_{r_{\rm hor}}^{r_{\rm max}} Ze E(r,a)dr,$$
where $H={r_{\rm hor}}-{r_{\rm max}}$.

By making use of  Eq.~(\ref{eq:1}) one obtains
$$
{\cal E}_{\rm max}(a)=2ZeMB_{0}\left.\frac{ra}{r^2+a^2}\right|_{r_{\rm hor}}^{r_{\rm
max}}
$$
\begin{equation}
\label{eq:4}
=2aZeMB_{0} \left[
\frac{(r_{\rm hor}+H)}{(r_{\rm hor}+H)^2+a^2}-\frac{r_{\rm hor}}{r_{\rm
hor}^2+a^2}\right].
\end{equation}
It is easy to see that the expression in square brackets equals to
$-H(H+2Mr_{\rm hor}-2a^2)$ and therefore is always negative
for any value of $H$ (remember that the angular momentum per unit mass $a$
varies between zero and $M$).
Thus for parallel magnetic field and angular momentum, $aM>0$ and the
radial component of the electric field $E_{\widehat{r}}$ on the black-hole
rotation axis is negative. So if there are negative charges near the
rotation axis, they will be accelerated away from the black hole.
According to Eq.~(\ref{eq:4}) the difference between their energies at
$r_{\rm max}$ and $r_{hor}$ is positive, so they gain energy while moving
along the rotation axis away from the black hole.
In the opposite case, when the magnetic field and angular momentum are
antiparallel, the radial component of the electric field on the black hole
rotation axis is positive.  So positive-charge particles, situated near the
rotation axis, are accelerated away from the black hole; according to
Eq.~(\ref{eq:4}), the energy gain in this
case is positive for the positive-charge particles.

\begin{figure}
\centering \includegraphics[width=0.48\textwidth]{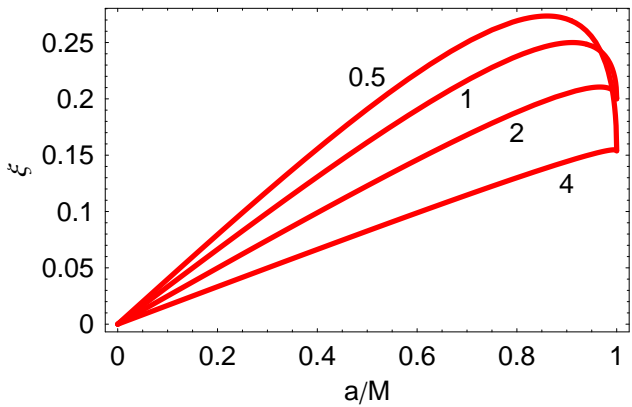}
\caption{\label{f_1}
Dependence of the normalized maximal energy $\xi$ from the angular
momentum $a$ of the black hole with mass $M$ for different values of the
vacuum-gap size $H$ (curves labeled by values of $H/M$; see the text for
definitions of parameters). }
\end{figure}

In Ref.~\cite{NerSemTk},
a simple expression ${\cal E}_{\rm max}\sim \  ZeB_{0}H$
rather than Eq.~(\ref{eq:4}) was used; however,
more precisely,
Eq.~(\ref{eq:4}) implies ${\cal E}_{\rm max}< \  ZeB_{0}H$.
The dependence of $\xi={\cal E}_{\rm max}/ZeB_{0}H$ from the angular
momentum $a$ for different $H$ is shown in Fig.~\ref{f_1}.
For black holes with angular momentum $a>0.1M$ and $H\sim
(1-2)R_{\rm S}=(2-4)M$, we have
${\cal E}_{\rm max}\sim 0.1\  ZeB_{0}H$; however, for slowly rotating
black holes, $a<0.1M$, $\ {\cal E}_{\rm max}$ varies from zero to
$\ \sim 0.01 ZeB_{0}H$. Hereafter, $R_{\rm S}$ denotes the Schwarzschild
radius.

However, the precise value of ${\cal E}_{\rm max}$ is often
irrelevant since particles cannot achieve this maximal energy because of
inevitable energy losses associated with the accelerated motion of the
particle. The particle energy is determined by the balance between the
energy losses and the energy gain per unit time,
$$\frac{d{\cal E}_{+}}{dt}=\frac{d{\cal E}_{-}}{dt}.$$
It was shown in Ref.~\cite{NerSemTk} that protons can be accelerated to
the energies of about $10^{20}$~eV only if the magnetic field is almost
aligned with the rotation axis.
In this case, only the curvature radiation is relevant for an accelerated
particle, and not the synchrotron one. We should also note here that we
do not consider energy losses related to interactions of accelerated
particles in the source. The maximal energy of an accelerated particle
is
$${\cal E}_{\rm curv}=\left(\frac{3}{2}\right)^{1/4}\frac{A}{Z^{1/4}}\
\frac{m}{e^{1/4}}\ E^{1/4} \ R^{1/2},$$
\begin{equation}
{\cal E}_{\rm curv} \approx 1.23 \times 10^{22}~\mathrm{eV} \ \frac{A}{Z^{1/4}}\
\left(\frac{B_0}{1~\mathrm{G}}\right)^{1/4}
\left(\frac{R}{1~\mathrm{kpc}}\right)^{1/2}{{\kappa}^{1/4}},
\label{Eq:Ecurv}
\end{equation}
where $B_{0}$ is the external magnetic field, $\ R\sim R_{\rm S}/\chi$  is
the curvature radius of magnetic-field lines, $Ze$ is the particle charge,
$Am$ is the particle mass ($A$ is the atomic number and $m$ is the nucleon
mass) and $\kappa$ is a coefficient between the electric field and
the external magnetic field, $|E_{\widehat{r}}|=\kappa B_{0}$ (see
Eq.~(\ref{eq:1})). Note that $\kappa$ is a function of the angular
momentum $a$ and the coordinate $r$. We point out that numerically,
${\cal E}_{\rm curv}<{\cal E}_{\rm max}$, cf.\ Ref.~\cite{UFN-HillasPlot},
and therefore it is ${\cal E}_{\rm curv}$ which determines the actual
maximal energy. For simplicity, we assume that all of the particles start
with equal initial conditions and so all of them are accelerated to the
same energy ${\cal E}_{\rm curv}$.

The original model of Ref.~\cite{NerSemTk} treated the magnetic field
$B_{0}$ as a free parameter. However, one may note that the field is
constrained and, in particular, cannot be too high (see
Ref.~\cite{UFN-HillasPlot} for a detailed discussion).
The maximal value of the magnetic field is determined \cite{Znajeck,
Abramovich} by the so-called Eddington limit,
$B_{\rm Ed}=10^{4}\left(\frac{M}{10^9 M_{\bigodot}}\right)^{-1/2}
\text{G}$. Though this estimate may be obtained in several different ways,
its most transparent meaning is that the magnetic-field energy density is
equal to that of the accreting plasma, corresponding to the Eddington
luminosity. To obtain the maximal possible particle energy, we should
assume that the external magnetic field is equal to the Eddington limit,
$B_{0}=B_{\rm Ed}$.
However, below, we will find the spectrum of cosmic rays in the frameworks
of this model. For this purpose we are interested in the actual maximal
particle energy as a function of black-hole mass $M$ rather than in the
upper limit. So we should recognize that the realistic magnetic field can
differ from the Eddington limit. In a general case, we can parametrize the
external magnetic field as follows:
$$B_{0}=k B_{\rm Ed}
\left(
\frac{M}{10^{9}M_{\bigodot}}
\right)^{\alpha},$$
where $\alpha$ and $k$ are some parameters. We note in passing that
several realistic models predict this kind of dependence, e.g.\ the
Shakura--Syunyaev model \cite{ShakuraSyunyaev, NovikovThorne} ($k\approx
0.31$, $\alpha=0$) or the model of Ref.~\cite{OtherBmodel} ($k\approx
0.0093$, $\alpha \approx -0.31$; see Fig.~2 of Ref.~\cite{UFN-HillasPlot}
for comparison with scarce observational data). We will consider these
options below.
We have
\begin{equation}
{\cal E}_{curv} \approx 2.9\times 10^{20}~\mathrm{eV} \ \frac{A}{Z^{1/4}}\!
\left(\frac{M}{10^{9}M_{\bigodot}}\right)^{\frac{3}{8}+\frac{\alpha}{4}}\!
{\left(\frac{\chi}{1^{\circ}}\right)}^{-\frac{1}{2}}\!(k\kappa)^{\frac{1}{4}}.
\label{Eq:EfromMass}
\end{equation}

During acceleration, the particle emits curvature photons. In what
follows, we will need to obtain an upper bound on this emission. The peak
energy of the photons is determined by the particle energy ${\cal E}$,
$$
{\cal E}_{\gamma}=\frac{3}{2}\frac{{\cal E}^3}{m^3 R},
$$
and for the upper limit, we take ${\cal E}={\cal E}_{\rm curv}$,
$$
{\cal E_{\gamma}}\sim 14~\mbox{TeV}~\frac{A^3}{Z^{3/4}}\left(\frac{M}{10^9
M_{\bigodot}}\right)^{\frac{1}{8}+\frac{3\alpha}{4}}\left(\frac{\chi}{1^{\circ}}\right)^{-1/2}
(k\kappa)^\frac{3}{4}.
$$

The ratio of luminosities in photons, $L_{\gamma }$, and in cosmic rays,
$L_{\rm CR}$, may be estimated by comparing the total available potential
difference in the acceleration region along the rotation axis to its
fraction, spent on the particle acceleration:
$$
\eta=\frac{L_{\gamma}}{L_{\rm CR}}=\frac{{\cal
E}_{\rm max}}{{\cal E}_{\rm curv}},
$$

\begin{equation}
\eta=3.12
\left(\frac{M}{10^9M_{\bigodot}}\right)^{\frac{1}{8}+\frac{3\alpha}{4}}
\left(\frac{\chi}{1^{\circ}}\right)^{1/2}
\frac{Z^{5/4}}{A}
\xi\kappa^{-1/4}k^{3/4},
\label{Eq:eta}
\end{equation}
where $(\xi\kappa^{-1/4})\sim (0.1-2)\ $ for $\ 0<a<M\ $ and $\
R_{\rm S}<H<6R_{\rm S}$. In numerical calculations presented below, we use
$\kappa^{1/4}=0.7$ and $\xi=0.25$, cf. Figs.~\ref{f_1},~\ref{f_2}.

To summarize, the model we use assumes a monochromatic spectrum  of
accelerated particles with $\mathcal{ E}=\mathcal{ E}_{\rm curv}$,
Eq.~(\ref{Eq:Ecurv}), in each particular source. The value of $\mathcal{
E}_{\rm curv}$ depends, within the assumed magnetic-field model, on the
SMBH mass $M$ only (in what follows, we do not consider acceleration of
other particles than protons). The overall flux from the source remains a
free parameter.

\section{Population of the sources and the observed spectrum}
\label{sec:population}
Having discussed the model of particle acceleration in a particular
source, we switch now to the population of sources. As we have seen, the
properties of a single source are determined, within the magnetic-field
model we choose to study, by the SMBH mass (the dependence from the spin
is weak). To reconstruct the UHECR spectrum one has to consider the
population of SMBHs distributed in mass and luminosity. For simplicity we
will assume that the \textit{mean} SMBH luminosity in cosmic rays is
related to its mass,
\begin{equation}
L_{\rm CR} \propto M^{\beta},
\label{Eq:Lcr}
\end{equation}
where $\beta$ is an additional model parameter.
Note that
not every black hole can work as a source, because the
source should possess some special properties (for example, small
inclination angle, the vacuum gap larger than $R_{\rm S}$,
 absence of  numerous charged particles in the vicinity of the black hole
which might imply a thin or even absent
accretion disc). The fraction of sources where the mechanism works is also
encoded in the mean luminosity, Eq.~(\ref{Eq:Lcr}). The observed
spectrum can be obtained by convolving the SMBH mass function with the
(monochromatic) single-source spectrum and Eq.~(\ref{Eq:Lcr}) and taking
into account the propagation effects.

\subsection{The SMBH mass function}
\label{sec:pop:mass-function}
Since dynamical measurements of SMBH masses are available for a very
limited number of cases only, it is a difficult task to find the mass
function precisely. Fortunately, a number of indirect methods to estimate
the SMBH mass are available (see e.g.\ Ref.~\cite{FF} for a review).
Despite having large uncertainties in individual measurements, these
methods are suitable for obtaining average characteristics of the SMBH
population, see e.g.\ Ref.~\cite{MassFunctionReview} for a review.

An important feature of the SMBH mass function is its evolution: SMBHs
grow fast, cf.\ e.g.\ Ref.~\cite{MassFunctionGrow}. For our calculation,
we use one of the most recent published redshift-dependent mass functions
\cite{MassFunctionWeUse}. Of two functions presented there, we choose to
use the one based on the stellar mass functions because it has smaller
statistical uncertainties. The systematic uncertainties of the mass
function may be judged from Ref.~\cite{MassFunctionReview} and are well
within the overall precision of our toy model.

\subsection{The observed spectrum}
\label{sec:pop:spectrum}
Before reaching the Earth, the accelerated protons may interact with
the cosmic microwave background. The main two processes modifying the shape
of the propagated cosmic-ray spectrum are photopion production and $e^+
e^-$-pair production. The former leads to a strong suppression of
the proton flux above few tens of EeV known as the GZK effect~\cite{G,
ZK},
why the latter mostly dominates the attenuation below 10 EeV leading to
the so-called ``dip'' feature in the spectrum~\cite{Hill:1983mk,
Berezinsky:1988wi}. We use the numerical code developed in
Refs.~\cite{propagation}. The code also traces secondary particles
produced
in the interactions. It makes use of the kinetic-equation approach and
calculates the propagation of  nucleons, stable leptons  and photons using
the standard dominant processes (see e.g. Ref.~\cite{reviewsPropag}).
\begin{figure}
\centering
\includegraphics[width=0.7\columnwidth,clip=true,angle=270]{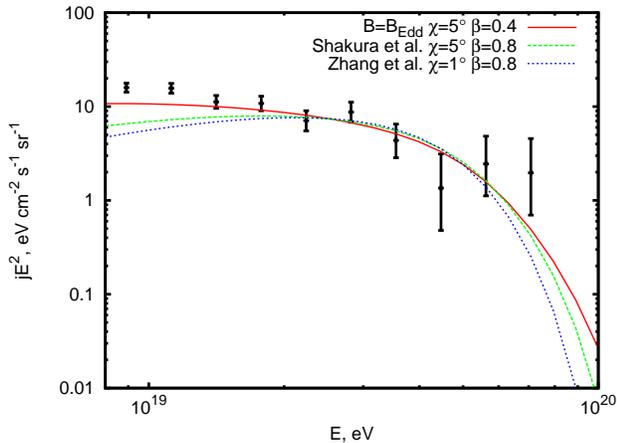}
\caption{The cosmic-ray flux predicted by the model for three different
assumptions about the SMBH magnetic field, see labels on the plot and
explanations in the text, versus the Auger experimental
data~\cite{AugerSpectrum}.}
\label{f_SpecAuger}
\end{figure}

Fig.~\ref{f_SpecAuger}
presents the predicted cosmic-ray
fluxes in the best-fit model for the
Auger spectrum \cite{AugerSpectrum} for different dependencies of
the magnetic field $B_0$ on SMBH mass mentioned in previous section. The
red curves correspond to $B_0$ given by the Eddington limit, the green one
describes the Shakura--Syunyaev model~\cite{ShakuraSyunyaev,
NovikovThorne}
and the blue one corresponds to the model of Ref.~\cite{OtherBmodel}. The
overall flux normalization is a free parameter. Besides we tried two values
of angle $\chi=1^{\circ}$ and $5^{\circ}$ and varied the
luminosity dependence~(\ref{Eq:Lcr}) parameter $\beta$ in the range
$-1<\beta<2$. The best-fit parameter values are indicated on the plot.
One can see that the first two models produce satisfactory spectral fits
above 10~EeV.

\section{Constraints}
\label{sec:constraints}
In this section, we discuss additional consistency
checks of the model. They include estimates of the accompanying gamma
radiation which should not be in conflict with the measured diffuse
gamma-ray background, estimates of the concentration of sources and of the
luminosity of a single source. We will see that the model passes these
tests. For order-of-magnitude estimates in this section, we assume
$B_{0}\sim B_{\rm Ed}$.

\subsection{Concentration of the sources}
\label{sec:constraints:density}
Let us check that the local concentration
of sources of cosmic rays with energies ${\cal E}\gtrsim  6\times
10^{19}~\mathrm{eV}$  is not in conflict with the lower
limit~\cite{PAO-clustering} based on the statistics of clustering. To this
end, we integrate the SMBH mass function over the range of masses
corresponding to these energies.

\begin{figure}
\centering \includegraphics[width=0.48\textwidth]{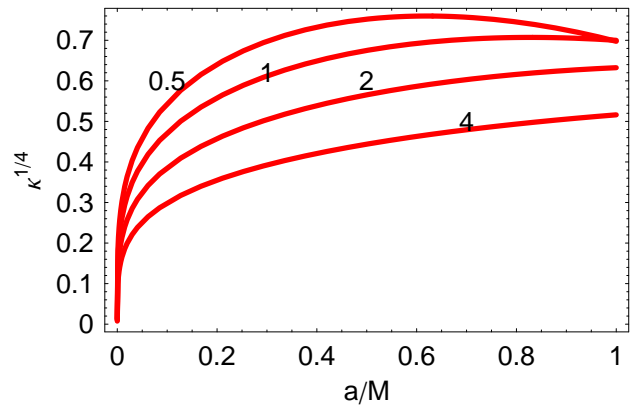}
\caption{\label{f_2}
Dependence of the coefficient $\kappa^{1/4}$ from the angular
momentum $a$ of the black hole with mass $M$ for different values of the
vacuum-gap size $H$ (curves labeled by values of $H/M$; see the text for
definitions of parameters).
}
\end{figure}

The dependence of the particle energy from the black-hole mass is given by
Eq.~(\ref{Eq:EfromMass}). Every black-hole mass corresponds to a range of
particle energies due to the variations in the value of $\kappa$. The
black-hole mass function includes all the black holes with fixed masses
and so all the black holes with every value of $\kappa$. Thus, the lower
limit of the required mass interval is determined by ${\cal E}= 6\times
10^{19}~\mathrm{eV}$ and the maximum value of $\kappa$. It was shown  in
Ref.~\cite{NerSemTk} that particles can be accelerated to the energies of
about $10^{20}~\mathrm{eV}$ only if the size of the vacuum gap is not
smaller than the Schwarzschild radius. In this case, the highest value of
$\kappa^{1/4}$ is $\sim 0.7$ (see Fig.\ref{f_2}).
Black-hole masses corresponding to particle energies  ${\cal E}\gtrsim
6\times 10^{19}~\mathrm{eV}$ are then $M\gtrsim 10^7 M_{\bigodot}$.
Integrating the mass function  in this range of masses, we obtain
$$
n=\int_{M_{\rm min}}^{M_{\rm max}}\frac{dn}{d\log M}d\log M.
$$
The
integral is saturated at its lower limit that is $M_{\rm max}$ can be
taken arbitrary high to obtain the following estimate,
\begin{equation}
\label{3}
n\sim  10^{-3}\ \frac{1}{\mathrm{Mpc}^3}.
\end{equation}
As we see, the total concentration of sources is larger than the
clustering lower bound of $10^{-4}\frac{1}{\mathrm{Mpc}^3}$. As we have
discussed above, only a
fraction of the calculated concentration $n$ corresponds to the true
concentration of the sources. Our estimate tells us that this fraction
should be not less than a few per cent which is reasonable.

\subsection{Luminosity of a single source}
\label{sec:constraints:luminosity}
A simple estimate of the luminosity of a single source may be obtained as
follows. Consider the observed
flux of cosmic rays with energies ${\cal E}\gtrsim  6\times
10^{19}~\mathrm{eV}$.  The value $j({\cal E})$, which is often reported, is
\begin{equation}
j({\cal E})=
\frac{1}{4\pi} \frac{1}{{\cal
E}}\frac{dF}{d{\cal E}}=\frac{1}{4\pi}\frac{dN}{d{\cal E}},
\label{Eq:j}
\end{equation}
where $N({\cal E})$ is the number of particles with energy $\leq {\cal E}
$, per unit area per unit time. Then, for the flux $F$ we have
$$
{\cal E}\cdot j({\cal E})=\frac{1}{4\pi}\frac{dF}{d{\cal E}},
$$
and, from the recent data~\cite{PAO-GZK, TA-GZK},
we obtain the estimate $F\sim 1.3 \times 10^5 \frac{\rm eV}{\rm m^2\cdot
s}. $

On the other hand, this flux is produced by the sources situated in the
GZK sphere, where the GZK horizon radius is $R_{\rm GZK}\sim 130\
 \mathrm{Mpc}$ for energies ${\cal E}\gtrsim  6\times 10^{19}~\mathrm{eV}$,
 see e.g.~\cite{GZKhorizon}.
We assume that all these sources have
approximately  equal cosmic-ray luminosities $L_0 \left[\frac{\rm eV}{\rm
s}\right]$, independent from the source black-hole mass (in reasonable
agreement with the best-fit values of $\beta$, Sec.~\ref{sec:population}).
Because in this case the GZK radius corresponds to very small redshifts
$z<0.1$, we can neglect changing of the source concentration with the
redshift. For the order-of-magnitude estimate, we neglect also the
difference between the energy with which the particle was emitted and the
final particle's energy with which we detect it on the Earth. Let us also
note that, because we are interested in all sources of the accelerated
 particles with energies ${\cal E}\geq 6\times 10^{19} \mathrm{eV}$, in
the expression for the total flux we have to substitute the total
concentration of sources, Eq.~(\ref{3}), that was obtained by integrating
the mass function.

The flux from every single source, situated at the distance $d$ from us,
is $F_{0}=L_{0}/(4\pi d^2)$. Thus for the total flux we have
$$
F=
\int_0^{R_{\rm GZK}} F_{0}\ n \ dV=
L_{0}\cdot  n \cdot R_{\rm GZK}.
$$
Using
 $
 n\sim 10^{-3}\
 \frac{1}{\mathrm{Mpc}^3}$,
 we obtain
$
L_{0}\sim
6\times 10^{39}\
\frac{\mathrm{erg}}{\mathrm{s}}.
$
The corresponding luminosity in photons  is
$
L_{\gamma}=\eta\cdot L_{0},
$
where $\eta$ is given by Eq.~(\ref{Eq:eta}).
One has
\begin{equation}
\label{4}
L_{\gamma}\sim  10^{40}\
\left(\frac{M}{10^{9}M_{\bigodot}}\right)^{1/8}
{\left(\frac{\chi}{1^{\circ}}\right)}^{1/2}\
\frac{\mathrm{erg}}{\mathrm{s}} .
\end{equation}
This value is much smaller than the typical bolometric luminosity of an
AGN,
$L_{\rm AGN}\sim \left( 10^{41}-10^{43} \right) \
\frac{\mathrm{erg}}{\mathrm{s}}$,
and by far does not exceed the Eddington limit,
 $ L_{\rm Ed}=10^{47} \ \left(\frac{M}{10^{9}M_{\bigodot}}\right)\
\frac{\mathrm{erg}}{\mathrm{s}}$.
Taking the concentration  of the sources of order of the lower limit,
$n\sim 10^{-4}\frac{1}{\mathrm{Mpc}^3}$, does not result in a conflict as
well.

In all the cases, the luminosity is not that far
from the luminosity of an AGN, and a natural question
arises: can we see our sources in $\mathrm{TeV}$ as  point sources? A
simple estimate of the flux of $\mathrm{TeV}$ photons associated with
the particle acceleration from one of the nearest sources, $R\sim
10~\mathrm{Mpc}$, gives the answer: we cannot see them as point sources
because the flux from a single source is smaller than the sensitivity of
the telescopes. Indeed, the flux of $ F_{\gamma}\sim 10^{-1}
\frac{\rm eV}{\mathrm{cm}^2\cdot\mathrm{s}} $ corresponds to the counting
rate of $ \sim 10^{-13}\frac{1}{\mathrm{cm}^2\cdot\mathrm{s}}, $ beyond
the reach of current TeV telescopes. Of course, this does not mean that
strong TeV sources cannot accelerate UHE particles by this mechanism: the
TeV emission may have a totally different origin.

\subsection{Diffuse gamma-ray background}
\label{sec:constraints:diffuse}
While individual cosmic-ray sources have quite low gamma-ray luminosities,
one may wonder about the total emission of all sources in the Universe
(beyond the GZK sphere). The emitted curvature photons have energies of
order a few TeV and interact with the infrared background radiation to
produce electromagnetic cascades in which the energy of the leading gamma
rays downgrade to the GeV band. Electrons in the cascade are deflected by
cosmic magnetic fields so distant sources contribute to the diffuse
gamma-ray background. Let us check that this contribution does not exceed
the measured value of the diffuse flux.

A simple estimate may be obtained as follows.
Consider particles with energies
${\cal E} \sim 10^{20}\mathrm{eV}$.
Cosmic rays with these energies arrive to us
from the interior of the GZK sphere only, but
the associated
photons come from all sources at all distances. Knowing how
much cosmic rays come from the interior of the GZK sphere, we can estimate
how much of them are present in the Universe (keeping in mind that the
number of sources depends on the distance). And supposing that the
luminosity of a source in cosmic rays is connected with its luminosity in
associated photons, Eq.~(\ref{Eq:eta}), we can estimate the total photon
emission from all the sources.

The flux at the Earth, Eq.~(\ref{Eq:j}), is expressed in terms of the
energy at detection, $\cal E$, which in general differs from the energy at
injection, $E_{\rm in}$. In particular, the account of the Universe
expansion (even neglecting additional energy losses) results in ${\cal
E}=E_{\rm in}(1+z)^{-1}$, where
$z$ is the redshift of the source.
The number of the emitted particles per unit time is also $(1+z)$ times
higher than the number of detected particles on the Earth per unit time.

The contribution to $j(E)$ from the sources located at redshift
$z$ is
$$
dj(E,z)=\frac{1}{4\pi}\ \frac{dn_{\rm BH}(M(E_{\rm in}))}{dE_{\rm in}} \
\frac{dN_{0}}{dt}\ \frac{1}{S(z)}\,dV(z),
$$
where $n_{\rm BH}(M(E_{\rm in}))$ is the number density of black holes with
masses $\leq M(E_{\rm in})$ (for $M(E)$, see Eq.~(\ref{Eq:EfromMass})) at
the redshift $z$, $dN_{0}/dt$ is a number of detected particles per unit
time from one of the sources, which were emitted with the energy $E_{\rm
in}$, $dV$ is the volume of a spherical layer at the distance $z$ from us,
$S$ is the area of a sphere with the radius equal to the distance from the
source to the Earth (it is necessary for calculating a number of particles
through the unit area on the Earth). Taking into account the dependence
from the redshift $z$, we may find the source cosmic-ray luminosity
$L(E_{\rm in})$, that is  its total energy emission in cosmic rays per
unit time,
$$
L(E_{\rm in})=\frac{dE_{\rm tot}}{dt}=E_{\rm in}\frac{dN_{0}^{\rm in}}{dt},
$$
where
$dN_{0}^{\rm in}/dt$ is a number of particles with the energy $E_{\rm in}$ emitted
by the source per unit time
and we assumed
that the source emits particles with
the only energy, determined by its mass $M$.

For the flat Universe, we have
$
S=4\pi a_0^2
\rho^{2}(z)$ and $ dV=Sa_{0}d\rho(z),
$
where
$$
 \rho(z)=\int_0^z\frac{dz}{H(z)a_{0}}
 $$
 is the geodesic coordinate distance
 from the
observer to the source location.
Finally
\begin{widetext}
 \begin{equation}
\label{f_9}
j(E)=
\int_0^{z_{\rm max}} \!
\frac{1}{4\pi}\frac{dn_{\rm BH}(M((1+z)E))}{d\log
\frac{M}{M_{\bigodot}}} \frac{L((1+z)E)}{(1+z)^{2}E}\ \left.
\frac{d\log \frac{M}{M_{\bigodot}}}{dE_{\rm in}}\right|_{M((1+z)E))}
 a_{0}d\rho(z),
\end{equation}
\end{widetext}
where $dn_{\rm BH}(M)/d\log M$ is just the mass function for a given
$z$~\cite{MassFunctionWeUse},
$z_{\rm max}$ is the redshift of the most
distant source.
Here we are interested in the values of the mass function at the points
$M((1+z)E)$ as a function of z. This function could be easily constructed
using the data from Ref.~\cite{MassFunctionWeUse}. Using
Eq.~(\ref{Eq:EfromMass}), we have
$$
\frac{d\log
\frac{M}{M_{\bigodot}}}{dE_{\rm in}}(M((1+z)E))
\sim \frac{1.16}{E_{\rm in}}=\frac{1.16}{(1+z)E}.
$$

For simplicity, let us suppose  that all the sources have the same
luminosities $L((1+z)E)=L_{0}$. This assumption is in reasonable agreement
with the results obtained in section~\ref{sec:population}. Thus we can
take the luminosity out of the integral. Let us now imagine just for a
moment, that cosmic rays with such energies could come from all distances,
and calculate the ratio of the values $j(E)$ for the cosmic rays with the
energies $\sim 10^{20}$~eV from the GZK sphere ($z_{\rm max}=z_{\rm
GZK}\sim 0.01$) and from the whole Universe (e.g.\ $z_{\rm max}=2$). The
luminosities in front of these two integrals are cancelled and after
integrating we obtain:
$$
\frac{E^2 j(E)_{\rm tot}}{E^2
j(E)_{\rm GZK}}=20.
$$
Taking into account the observed value of the cosmic-ray flux, we estimate
$$
E^2 j(E)^{\rm tot}\sim 10^5 \frac{\mathrm{eV}}{\mathrm{m}^2
\cdot\mathrm{s}\cdot \mathrm{sr}}.
$$

For calculating the value of $E_{\gamma}^2 j(E)_{\gamma}$ for the gamma
radiation from all the possible sources in the whole Universe we should
only replace $d\log M/dE$ in the integral (\ref{f_9}) by the
corresponding expression for photons, Eq.~(\ref{4}),
and the cosmic-ray luminosity $L$ by the photon luminosity $L_{\gamma}=\eta
L$, Eq.~(\ref{Eq:eta}). The order-of-magnitude estimate then reads
$$
E_{\gamma}^2
j(E)_{\gamma}^{\rm tot}=\eta E^2 j(E)^{\rm tot}\sim
10^5\frac{\mathrm{eV}}{\mathrm{m}^2 \cdot\mathrm{s}\cdot \mathrm{sr}}.
$$
This total gamma-ray emission associated
with particle acceleration in all the sources does not exceed the observed
value of the diffuse gamma-ray background \cite{FERMIdiffuse}.

\begin{figure}
\centering
\includegraphics[width=0.7\columnwidth,clip=true,angle=270]{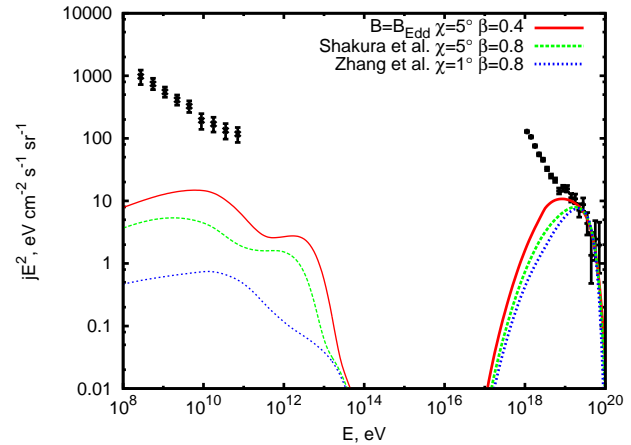}
\caption{Gamma-ray fluxes predicted by the same models as shown in
Fig.~\ref{f_SpecAuger}.}
\label{f_SpecAugerGamma}
\end{figure}

We have also performed a more detailed numerical simulation of the
secondary gamma-ray flux. The injection spectrum of the curvature photons
is similar to the synchrotron one ~\cite{Electrod},

$$
I(\omega)\approx 2\sqrt{3} Z^2 e^2 \gamma
\frac{\omega}{\omega_{c}}\int_{2\omega/\omega_c}^{\infty}K_{5/3}(x)dx,
$$
where
$$
\omega_{c}= 3 \frac{\gamma^3}{R}
$$
is the ``critical'' frequency
(for higher frequencies, radiation is negligible), $\omega$ is the
frequency of the radiated photons, $K_{5/3}$ is the Macdonald function,
$\gamma=({\cal E}/m)$ is the particle Lorentz factor, $R$ is the curvature
radius of the particle trajectory (for curvature radiation, $R$ is
constant and it is equal to the curvature radius of the magnetic lines).
The total flux of the photons from a single source is related to the
cosmic-ray flux by the coefficient $\eta$, Eq.~(\ref{Eq:eta}). For the
best-fit spectrum of Sec.~\ref{sec:pop:spectrum}, we used the same code to
describe the propagation of the accompanying gamma rays and to calculate
the observed gamma-ray flux. The result is presented in
Fig.~\ref{f_SpecAugerGamma}.
One may see that the diffuse gamma-ray upper limit is satisfied.

\section{Conclusions and discussion}
\label{sec:concl}
We have constructed and studied a toy model of UHECR acceleration in the
vicinity of numerous and various supermassive black holes in centers of
galaxies. The model assumes that:
\begin{itemize}
 \item
cosmic-ray particles are accelerated by the regular electric field within
a few $R_{S}$ from the SMBH~\cite{NerSemTk}; the field configuration is
given by the solutions of Refs.~\cite{BHsolution1, BHsolution2} and is
fully determined by the SMBH mass $M$, its angular momentum $a$ and the
magnetic-field normalization $B_{0}$;
\item
all cosmic-ray particles accelerated near a given SMBH have similar
initial conditions and therefore all are accelerated up to one and the
same energy limited by the curvature-radiation losses; this maximal energy
is calculated in the model and depends on $M$ only (provided $B_{0}$ is a
given, model-dependent, function of $M$; dependence from $a$ smooths this
monochromatic spectrum insignificantly);
\item
the \textit{mean} flux of a source (which accounts for the fraction of the
sources where this mechanism does work) depends from $M$ in a power-like
manner; the normalization and the exponent are two free parameters of the
model (the best fit to the cosmic-ray spectrum indicates that this
dependence is weak);
\item
the concentration of sources is determined by the redshift-dependent SMBH
mass function taken from astrophysical literature.
\end{itemize}
Within these assumptions and given the $B_{0}(M)$ relation is fixed (we
considered three popular choices for it), the model has two free
parameters which we find by fitting the cosmic-ray spectrum at the Earth
to the experimental data. With parameters fixed in this way, we subject
the model to several further tests which it passes succesfully:
\begin{enumerate}
 \item
the concentration of sources is large enough to satisfy the constraints
from absence of clustering in UHECR arrival directions;
\item
the luminosity of a particular source, determined by the flux
normalization and concentration, is not too high;
\item
secondary gamma rays from distant sources do not overshy the measured GeV
diffuse gamma-ray background.
\end{enumerate}
Given the success of the toy model, it is interesting to discuss its
possible refinements.
The assumptions we have made within the model are quite robust and
realistic. One subtle point is related to the value of the SMBH angular
momentum $a$ which may vary from one black hole to another. However, these
variations are probably modest given the scaling relation between the mass
and the angular momentum of cosmic black holes proposed in
Ref.~\cite{Nature-momentum}. The precision of predictions may be improved
with more realistic modelling of the acceleration mechanism, in
particular, with account of the charge concentration at the SMBH and in
the acceleration region, of the finite thickness of the accretion disk
etc. Ultimately, this approach might give answer to the question, which
particular SMBHs are strong sources and which are not, thus determining
the (presently free) parameter theoretically. However, this is a
complicated task and is far beyond the scope of the present work.

One possible question concerns the low-energy part of the spectrum where,
as is clearly seen from Fig.~\ref{f_SpecAugerGamma}, the contribution of
the mechanism we discuss is insufficient to explain the observed spectrum
due to the depletion of the SMBH mass function at low masses. It is
tempting to speculate that this depletion is compensated by a huge
contribution of the SMBH in our own Galaxy which, indeed, has the
appropriate mass. A quantitative analysis of this proposal requires,
however, a much more precise study of physical properties and
possibilities for particle acceleration close to the Galactic Center.

{\bf Acknowledgements. } We are indebted to V.~Alba, P.~Dunin-Barkowski,
G.~Farrar, A.~Neronov, D.~Semikoz and I.~Tkachev for interesting
discussions. We thank the authors of Ref.~\cite{MassFunctionWeUse} for
providing numerical values of the SMBH mass function we used. This work
was supported in part by the RFBR grant 10-02-01406 and by the grant of
the President of the Russian Federation NS-5590.2012.2, by the RFBR grants
12-02-01203 and 11-02-01528 (S.T.) and by the Dynasty Foundation (K.P.\
and S.T.). Numerical part of the work was performed at the cluster of the
Theoretical Division of INR RAS.

%%%%%%%%%%%%%%%%%%%%%%%%%%%%%%%%%%%%%%%%%%%%%%%%%%%%%%%%%%%%%%%%%%%%%%%%%%%

\end{document}